%% file: bare_jrnl.tex
\documentclass[conference]{IEEEtran}
\IEEEoverridecommandlockouts
\usepackage[numbers]{natbib}
\setcitestyle{open={[},close={]}}

\usepackage{amsmath,amssymb,amsfonts}

\usepackage{graphicx}
\usepackage{textcomp}
\usepackage{xcolor}
\usepackage{float}
\usepackage{array}
\usepackage[font=footnotesize,labelfont=bf]{caption}
\usepackage{optidef}
\usepackage{subcaption}
\usepackage[utf8]{inputenc}
\def\BibTeX{{\rm B\kern-.05em{\sc i\kern-.025em b}\kern-.08em
    T\kern-.1667em\lower.7ex\hbox{E}\kern-.125emX}}

\usepackage{algorithm}
\usepackage{algpseudocode}
\usepackage{color, xcolor, colortbl}

\begin{document}

\title{Data-Quality Based Scheduling for Federated Edge Learning\\}
\author{\IEEEauthorblockN{
		Afaf Taïk, \IEEEmembership{Student Member, IEEE}, Hajar Moudoud, \IEEEmembership{Student Member, IEEE} \\
    and Soumaya Cherkaoui, \IEEEmembership{Senior Member, IEEE} }	


 }

\maketitle
\begin{abstract}
\input{./0_abstract}
\end{abstract}

\IEEEpeerreviewmaketitle

\section{Introduction}
\label{sec: Introduction}
\input{./1_introduction}
\section{Related work}
\label{sec: Related works}
\input{./2_related}

\section{System Model}
\label{sec: system}
\input{./3_system_model}

\section{Data-Quality based Scheduling}
\label{sec:proposed}
\input{./4_proposed}
\section{Numerical results}
\label{sec: evaluation}
\input{./5_simulation}
\section{Limitations and Future Work}
\label{sec: limitations}
\input{./6_limitations}
\section{Conclusion}
\label{sec: Conclusion}
\input{./7_conclusion}
\section*{acknowledgement}
\label{sec:acknowledgement}
The authors would like to thank the Natural Sciences and Engineering Research Council of Canada, as well as FEDER and GrandEst Region in France, for the financial support of this research.

\label{Related work}
\bibliographystyle{IEEEtran}
\bibliography{./references}


\end{document}

%% file: 0_abstract.tex
FEderated Edge Learning (FEEL) has emerged as a leading technique for privacy-preserving distributed training in wireless edge networks, where edge devices collaboratively train machine learning (ML) models with the orchestration of a server. However, due to frequent communication, FEEL needs to be adapted to the limited communication bandwidth. Furthermore, the statistical heterogeneity of local datasets' distributions, and the uncertainty about the data quality pose important challenges to the training's convergence. Therefore, a meticulous selection of the participating devices and an analogous bandwidth allocation are necessary. In this paper, we propose a data-quality based scheduling (DQS) algorithm for FEEL. DQS prioritizes reliable devices with rich and diverse datasets. In this paper, we define the different components of the learning algorithm and the data-quality evaluation. Then, we formulate the device selection and the bandwidth allocation problem. Finally, we present our DQS algorithm for FEEL, and we evaluate it in different data poisoning scenarios.

%% file: 1_introduction.tex
Training centralized machine learning (ML) models requires the collection of large datasets from sparse and highly distributed sources. Nonetheless, in addition to the communication overheads, data collection raises an increasing concern about sharing private information. FEderated Edge Learning (FEEL) \cite{r8,tak_federated_2021} is a ML setting that utilizes multi-access edge computing (MEC) to tackle these concerns. In contrast to centralized ML, FEEL consists of training the model on the user equipment (UE) under the orchestration of a MEC server, where only the resulting parameters are sent to the edge servers to be aggregated.

The  predominant FEEL system model is a synchronous time-division resource allocation system \cite{bonawitz_towards_2019}. This means that the MEC server waits for all UEs to send their updates before aggregating them. However, in reality, not all updates can be aggregated. In fact, updates from UEs who do not report before a fixed deadline are cancelled. A real deployment of FEEL is therefore subject to the following challenges:   

 {\bf Limited Resources}:  In a contrast to cloud servers, the computing and storage resources at the edge are rather limited \cite{moudoud1,abouaomar_resource_2021}. The gap in computational resources of the UEs causes significant delays due to stragglers. Furthermore, to run the iterative FEEL learning, the Base Station (BS) generally needs to connect a large number of UEs across a resource-limited spectrum and therefore can only support a limited number of UEs sending their model updates over unreliable channels for global aggregation.  Additionally, the size of the updates can be very large in the case of deep neural networks. As a result, the communication overhead becomes a bottleneck for FEEL. 
 
 {\bf Unbalanced and non-IID}:  Unlike traditional ML systems, in which an algorithm operates on a large data set distributed homogeneously over several servers in the cloud, FEEL is typically trained on a large, often unbalanced, and non-identically distributed (non-IID) dataset that is generated by separate distributions across different UEs. A large number of UEs participate in the training, where several own small datasets, which makes local models prone to overfitting. Moreover, the dataset of a given UE is usually based on a particular user's usage pattern, and thus a particular user's local data set will likely be redundant and not representative of the remainder of the population distribution.  
 
 {\bf Unreliable and malicious devices}: Although FEEL has improved privacy protection, it still suffers from several problems. First, all devices involved in the FEEL process are expected to contribute their resources unconditionally, which is not accepted by all UEs. Without reward, only a small fraction of the devices are willing to participate in the training process. On the other hand, the UEs involved in the training are unreliable and may act maliciously, intentionally, or unintentionally, which may affect the overall model and lead to erroneous model updates. 

 Most existing work on FEEL proposed scheduling algorithms aiming to optimize resource utilization. The considered resources are time \cite{rh7}, transmission energy of participating UEs \cite{rh6}, and local computation energy \cite{rh5}. Consequently, the number of scheduled UEs is often restricted in order to meet latency and energy constraints. However, optimization should consider learning related aspects, especially data distributions and quality \cite{rh7,goetz_active_2019,taik_data-aware_2021}. Furthermore, providing a reliable device selection is a necessary stepping stone for enabling efficient and useful updates \cite{rh4}. 

The main motivation behind this paper is to design a scheduling algorithm for FEEL that considers UEs datasets' different properties at its heart. In fact, while datasets are unbalanced and non-IID, but a UE with a large dataset is not necessarily in possession of useful data. Furthermore, malicious UEs can train the model on poisoned datasets, which requires additional mechanisms to reduce their damages to the training process.
In this paper, we consider diversity of the datasets and the reliability of the UEs as the baseline criterion for choosing participating devices in FEEL. The diversity evaluation allows to give priority to UEs with potentially more informative datasets to speed up the training process, and the reputation mechanism sanctions malicious UEs with poisoned datasets. 
Accordingly, we propose a method for incorporating datasets' properties in FEEL scheduling, by designing a novel data-quality value measure, and designing a data-quality based scheduling algorithm (DQS).

The contributions of this paper can be summarized as follows:
\begin{itemize}
    \item[1)] we design a suitable priority indicator for the selection of participating UEs;
    \item[2)] we formulate a joint device selection and bandwidth allocation problem taking into account data properties; 
    \item[3)]we prove that the formulated problem is NP-hard and we propose a DQS scheduling algorithm based on greedy knapsack; and
    \item[4)] we evaluate the proposed data-quality value measure and the DQS algorithm through insightful simulations. 
\end{itemize}

The remainder of this paper is organized as follows. In Section II, we present the background for FEEL and related work. In Section III, we present the design of the proposed diversity measure, starting with the used uncertainty measures and their integration in FEEL. In Section IV, we integrate the proposed measure in the design of the joint selection and bandwidth allocation algorithm. Simulation results are presented in Section V. At last, conclusions and final remarks are presented in Section VI.

%% file: 2_related.tex
In this section, first we discuss the related work regarding device scheduling and data quality in FEEL. Next, we illustrate the existing research gaps and motivate the need for designing a data-quality based scheduling algorithm for FEEL.

Among the first works to explore device selection is work in \cite{fedcs}, where authors mitigated the straggler problem by giving priority to end devices with good communication and computation capabilities. However, this method is not suitable in FEEL context, where datasets' distributions vary, as it discards UEs with larger and potentially more informative datasets. Several works proposed scheduling algorithms that aim to maximize the number of participating devices while optimizing used resources \cite{rh8,rh9}. For instance, authors in \cite{zeng_energy-efficient_2019} proposed an energy-efficient scheduling algorithm aiming to collect the maximum number of updates possible as a guarantee for the training speed. Another example is in \cite{yang_age-based_2020}, where authors study different scheduling strategies based on the wireless channel's state. Other works focus on the staleness of updates to calculate the priority \cite{yang_age-based_2020,xia_federated-learning-based_2021}, where higher priority is given to UEs that did not participate in previous rounds or have stale updates. Despite the variety of the scheduling algorithms in the literature, the design and evaluation of FEEL scheduling  under heterogeneous and uncertain dataset distributions remains a topic that is not well addressed, as data-quality issues are overlooked and under-explored.

To overcome the reliability problem,  several FL works use reputation systems for UEs selection, as UEs with high reputation are more likely to provide high quality data for the learning procedure. For instance, the authors in \cite{rh1} proposed a decentralized peer-to-peer approach to overcome the problem of unreliable UEs, where poisoning is prevented by verifying peer contributions to the model. However, due to stastical heterogeneity and high communication costs, P2P verification may not always be suitable. In \cite{rh3} and \cite{rh2}, the authors presented a blockchain-based reputation system for FL, where a  UE's reputation is traceable. In \cite{rh2}, the authors presented a reputation-based UE selection using subjective logic. This solution evaluates the reputation of UEs based on their past interactions with other task publishers in the network. Nonetheless, such approach relies on the willingness of several other similar task publishers to share their opinions.  
Based on our analysis of these works, we found that most of them did not consider the data heterogeneity when evaluating device reputation, and might not be suitable for realistic FEEL scenarios.
Consequently, we found it necessary to build a scheduling algorithm for FEEL that considers UEs datasets’ different properties ($i.e.,$ limited bandwidth, unbalanced/ non-IID data) alongside UEs reliability.

%% file: 3_system_model.tex
\begin{figure}[t]
	\centering
	\includegraphics[width=8.5cm, height=6.2cm]{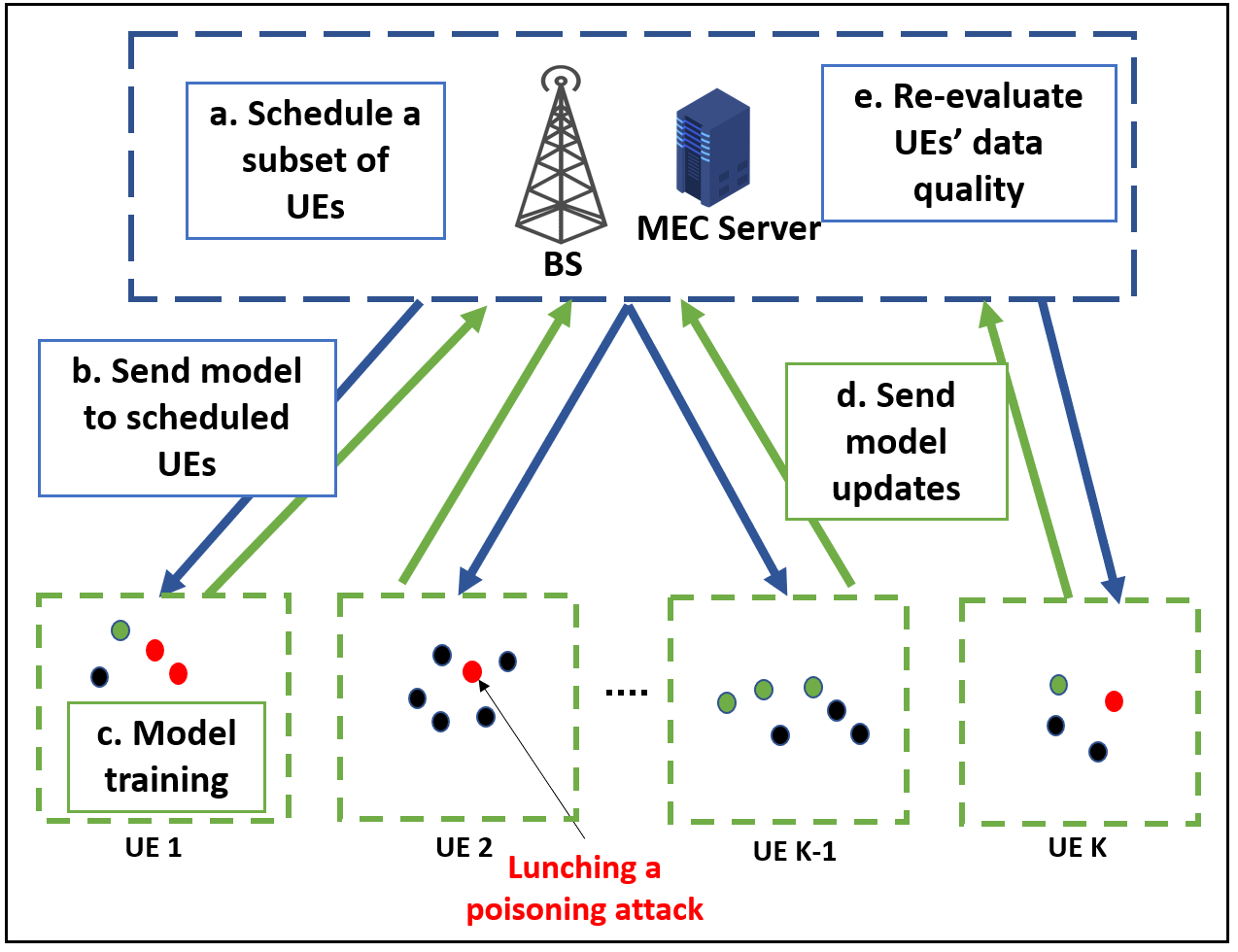}
	\caption{Illustration of a communication round of FEEL with $\it{K}$ UEs. U$E_{2}$ launches a poisoning attack which will affect the learning. }
	\label{fig:archi}
\end{figure}

In this section, we introduce the different components of the system model. The system's main components are the learning model, the data-quality model, and the communication model. Based on these components, we formulate a joint UE selection and bandwidth allocation problem for FEEL. 

\subsection{Learning Model}
In contrast to centralized training, FEEL keeps the training data locally, and learns a global model through the shared parameters sent by a federation of participants (e.g., individual users, organizations). By keeping data locally, the training can use ephemeral data and leverage the computing resources of the sparse UEs. 
FEEL is an iterative process that starts with the model and parameters initialization. In our proposed framework, in each round, the procedure in Algorithm \ref{alg:procedure} is repeated. 
Fig. \ref{fig:archi} illustrates the different steps in the procedure taking into account the statistical heterogeneity of the data distributions alongside the possibility that malicious devices may launch data poisoning attacks. In the figure, $UE_2$ launches a label-flipping attack. Due to the local nature of the training, this attack is hard to detect.
As a result, in our proposed process, each training round $\it{t}$ starts by the UEs sending their information to the MEC server (i.e., dataset information).
With the evaluated channel state information, and the dataset information, the MEC server schedules a set of UEs using Algorithm \ref{alg:schedule} and sends them the global model  $\it{g}$. Upon receiving the updates, the MEC server evaluates the quality of each model $\Omega_k$ by testing it on publicly available data. UEs also report local accuracy of their models. These evaluations are then used to update the perceived reputation of the UEs at the MEC server level. 
The last step is the model aggregation, which is typically achieved using weighted aggregation \cite{r8}. 

\subsection{Data Quality and Data Poisoning}
\vspace{0.2cm}
\textit{1) Data Poisoning in FEEL}
In FEEL, UE data may include personal information ($e.g.,$ home address, credit card number, etc.). The disclosure of this data is not only harmful to the UE, but the intentional/unintentional alteration of the data can also cause security problems \cite{moudoud2}. One of the main types of attacks that can affect model severally is poisoning attack. A poisoning attack occurs when the attacker is capable of injecting false data or altering the training samples of the FEEL model's learning pool, and thus causing it to learn on wrong or erroneous data. Poisoning attacks can occur during the training period and are primarily aimed at availability or integrity of the data. Generally, there are two main approaches to generate poisoned attacks, namely: label-flipping \cite{cao_understanding_2019} and backdoor \cite{bagdasaryan_how_2020}. In this paper, we mainly focus on label-flipping where an adversary modifies a small number of examples/training data and maintains the characteristics of the data unchanged to degrade the performance of the model.

\vspace{0.2cm}
\textit{2) Reputation evaluation:}
At each communication round $\it{t}$, each participating UE $k$ reports the local accuracy $acc_{k}^{local}$ and sends its newly computed model $\Omega_k$. The model is then evaluated on a test-set to evaluate its quality, alongside the UE's honesty. A UE's reputation decreases when it uploads a malicious or bad-quality model, or when it declares a very high local accuracy compared to the obtained accuracy on the test-set.
This reputation measure allows to detect not only malicious UEs, but also UEs with overfitting models and low quality dataset. 
We update the value of the reputation of UE $\it{k}$ in each round $\it{t}$, as follows:
\begin{equation}
    R_{k}^{t} =  R_{k}^{t-1} - \eta ( \beta_1 (acc_{k}^{local} -
    avg(acc)) + \beta_2 (acc_{k}^{local} - acc_{k}^{test}))       
\end{equation}
where $\eta \in [0.1]$ is the reputation rate with which we decrease the reputation value.

\vspace{0.2cm}
\textit{3) Dataset diversity evaluation:}

Each dataset can be characterized by how diverse its elements are (i.e., diversity of the elements), their number(i.e., dataset size) and how many times the model was trained on it (i.e., age). 
We set the value of each metric as \cite{taik_data-aware_2021}: $ v_i \gamma_i$,  where  $\gamma_i$ is the adjustable weight for each metric assigned by the server and $v_i$ is the normalized value of the metric $\it{i}$.
Using the aforementioned characteristics, the diversity index of dataset $k$ can be defined as: 
\begin{equation}
I_k = \sum_{i} v_{i,k}\gamma_{i},
\end{equation}
with $i \in\{\text{elements diversity}, \text{dataset size}, \text{age}\}$
Depending on the application, other quality measures can also be used. For instance, in image classification, measures regarding image quality (e.g., blur) can heavily affect the learning \cite{ye_federated_2020,lakoju_chatty_2021}. Such measures can be evaluated on a subset of images and normalized for a seamless integration in the index. 
\begin{table}[t]
	\centering
	\caption{List of Notations.}
\resizebox{\columnwidth}{!}{	
\begin{tabular}[h]{l|l}
	\hline
	\textbf{Notations}&\textbf{Description}\\
	\hline
	$T$& Communication round's deadline\\
	$B$& Bandwidth\\
	$s$& Model size\\
	$\epsilon$& Number or local epochs\\
	$g$ & Global model\\	
	$\Omega_k$ & Model update of UE $\it{k}$\\
	$t_{k}^{train}$ & Training time for UE $\it{k}$\\
	$r_k$ & Achievable data rate of UE $k$\\
	$t_{k}^{up}$ & Upload time for UE $\it{k}$\\	
	$P_k$ & Transmit power of UE $k$\\
	$\alpha_k$ & Bandwidth fraction allocated to UE $k$\\
	$N_0$& Power spectral density of the Gaussian noise\\
	$\left | D_k \right |$ & UE $k$'s dataset size\\
	$\zeta_k (cycles/bit)$ & The number of CPU cycles\\
	$f_k$  & The  computation capacity\\
	$R_{k}^{t}$& Reputation of UE $\it{k}$ at round $\it{t}$\\
	$I_{k}$& Dataset diversity index of UE $\it{k}$\\
	$V_{k}$& Data-quality value of UE $\it{k}$\\
	$\eta$& Reputation rate\\
	$\beta_i, \omega_i, \gamma_i$& Weights of different measures\\
	\hline
\end{tabular}}
	\label{tab:caption}
\end{table}%

\vspace{0.2cm}
\textit{4) Data-Quality measure}
The data-quality value can be inferred from the reputation of UE $\it{k}$'s reputation and dataset diversity index. We define the value for each UE as:
\begin{equation}
    V_{k} =  \omega_1  R_{k} + \omega_2 I_{k}       
\end{equation}
with $ \omega_1, \omega_2 \in [0, 1]$ are weighted values for each metric.

\subsection{Communication and Computation Models}
As communication is the bottleneck of synchronous FEEL, it is crucial to judiciously allocate the bandwidth.  
Hereinafter, we consider orthogonal frequency-division multiple access (OFDMA) for local model uploading from the UEs to the MEC server, with total available bandwidth of $\it{B}$ Hz. We define $\alpha = [\alpha_1,...,\alpha_K]$, where each UE $\it{k}$ is allocated a fraction $\alpha_{k} \in [0,1]$ from the total bandwidth $\it{B}$. Additionally, we denote the channel gain between UE $k$ and the BS by $h_{k}$. The achievable rate of UE $\it{k}$ when transmitting the model to the BS is given by: 
\begin{equation} \label{eq:eqRate}
r_k=\alpha_{k} B \log_{2}(1 +\frac{g_{k} P_{k}}{\alpha_k B N_0}),\qquad \forall k \in [1,K],
\end{equation}
where $P_{k}$ is the transmit power of UE $\it{k}$, and $N_{0}$ is the power spectral density of the Gaussian noise.

Based on the synchronous aggregation assumption with a fixed deadline, the scheduled UEs in a communication round must upload before a deadline $\it{T}$. 
For all UEs, the time constraint is defined as follows: 
\begin{equation} \label{eq:eqTime}
  (t_{k}^{train}+t_{k}^{up}) x_k \leq T ,\qquad   \forall k \in [1,K],
\end{equation}
where $t_{k}^{train}$ and $t_{k}^{up}$ are, respectively, the training time and upload time of UE $\it{k}$. The training time $t_{k}^{train}$ depends on UE $k$'s dataset size as well as on the model. It can be estimated using Eq.\ref{eq:ttrain}: 
\begin{equation}
    t_{k}^{train} = \epsilon \left | D_k \right | \frac{\zeta _k}{f_k},
    \label{eq:ttrain}
\end{equation}
where $\zeta_k$(cycles/bit) is the number of CPU cycles required for computing one sample data at UE $k$ and $f_k$  is its computation capacity.

In order to send an update of size $\it{s}$ within a transmission time of $t_{k}^{up}$, we must have:
\begin{equation}
t_{k}^{up} = \frac{s}{r_k}.    
\end{equation}
\subsection{Joint UE selection and bandwidth allocation problem}

Considering the data-quality aspect at the key selection criterion, and taking into account the communication bottleneck of FEEL, we formulate the following problem: 
\begin{maxi!}|c|[3]
 {x,\alpha}{{\sum_{k=1}^{K}{x_{k}V_{k}}}}
 {}{} \label{pbmoo}
\addConstraint{\qquad (t_{k}^{train}+t_{k}^{up}) x_k \leq T, \qquad \forall k \in [1,K]}\label{eq:TrainUploadT}
\addConstraint{\qquad \sum_{k=1}^{K}\alpha_k  \leq 1,\qquad  \forall k \in [1,K]}\label{eq:Bandwidth}
\addConstraint{\qquad  0\leq \alpha_k \leq 1,\qquad \forall k \in [1,K]}\label{eq:alpha_bounds}
\addConstraint{\qquad x_k \in \{0,1\}, \qquad \forall k \in [1,K].}\label{eq:x_bounds}
\end{maxi!}

The goal of the problem (8) is to select UEs to participate in a training round, and allocate the bandwidth to these UEs so as they upload their models before the deadline. The goal is therefore to maximize the weighted sum of the selected UEs under time and bandwidth constraints.  
In fact, constraint (\ref{eq:TrainUploadT}) guarantees that the selected UEs will finish training and uploading before the deadline $\it{T}$. Due to limited bandwidth budget, the bandwidth allocation ratio should respect the constraint (\ref{eq:Bandwidth}). Constraints (\ref{eq:alpha_bounds})  determines the bounds for the bandwidth allocation ratios and constraint (\ref{eq:x_bounds}) defines $x_k$ as a binary value for each UE.

Problem (8) is mixed integer non-linear problem, which makes it very challenging to solve. Indeed, a restricted version of the problem~(8) can be shown to be equivalent to a knapsack problem and thus it is NP-hard. In fact, the problem of maximizing the weighted number of devices, i.e., $\sum_{k}V_k x_k$ is subject to a knapsack capacity (i.e., limited bandwidth) given by $\sum_{k}\alpha_k x_k\leq1$. Hence, the problem is equivalent to a knapsack problem, and since the latter is NP-hard, so is the problem~(8).

\begin{algorithm}[h]
	\begin{algorithmic}[1]
	\For{$\it{t} \in [1 \dots t_{max}]$}
	    \If{$t=0$}
	        \State initialize the model's parameters at the MEC server
	        \State initialize the reputation values as 1 for each UE. 
	    \EndIf
		\State Receive UEs information (transmit power, available data size, dataset diversity index)
		\State Schedule a subset $S_t$  of UEs with at least $\it{N}$ UEs using Algorithm \ref{alg:schedule}
		\For {\text{ UE } $k \in S_t$ }
			 \State $\it{k}$ receives model $\it{g}$
			 \State $\it{k}$ trains on local data $D_{k}$ for $\epsilon$ epochs
			\State $\it{k}$ sends updated model $\Omega_k$ to MEC server and reports local accuracy $acc_{k}^{local}$
		\EndFor
		\State MEC server computes new global model using weighted average:
		 $g\leftarrow \sum_{k \in S_t }\frac{D_{k}}{D_t}\Omega_{k}$
		 \State MEC server updates the reputation values using the received models using a test-set and the average reported accuracies.
        \State start next round $\it{t}\leftarrow \it{t+1}$		
	\EndFor
	\end{algorithmic}
	\caption{Data-quality based training procedure}
	\label{alg:procedure}
\end{algorithm}

%% file: 4_proposed.tex
In this section, we present our proposed solution for data-quality based scheduling to solve (8). As problem (8) can be transformed into a knapsack problem, we chose to follow a greedy knapsack algorithm to solve it. A greedy knapsack algorithm has low complexity and will allow fast and efficient scheduling under the rapidly changing wireless edge environment. In fact, while the ranked list solution of the knapsack approach might enhance scheduling performance in terms of overall data quality, it is not efficient in defining the optimal bandwidth allocation decision.  Moreover, selecting the highest ranked UE with the best data-quality value without considering the extent of bandwidth required by each UE does not lead to an optimized performance benefit. Consequently, it is preferable to follow a greedy algorithm that iteratively selects a UE with better ratio $\frac{V_k}{\alpha_k}$ \cite{cormen_introduction_2009}.

As the required bandwidth is not a fixed value, we calculate the minimum required bandwidth for each UE to be able to send the update by the deadline $\it{T}$.
In order for a UE $\it{k}$ to upload the model before the deadline $\it{T}$, each selected UE must respect $t_{k}^{up} \leq T - t_{k}^{train}$. The corresponding minimum data rate is therefore $r_{k,min} = \frac{s}{T - t_{k}^{train}}$ for each UE $\it{k}$. 
Calculating the value of $\alpha_k$ from this expression is not possible. Thus, we estimate the cost of allocating a fraction of the bandwidth to UE, by calculating the required number of fractions needed if we were to uniformly allocate the bandwidth. More specifically, we define 
\begin{equation}
r_k(c) = \frac{c}{K}B \log_{2}(1 +\frac{g_{k} P_{k}}{\frac{c}{K} B N_0}) ,c \in \mathbb{N} \cap [1,K]     
\end{equation}

As a result, we define the cost for bandwidth allocation for UE $\it{k}$  $c_k = min \{ r_k(c)\geq r_{k,min} , c \in \mathbb{N}\cap [1,K]\}$. The cost is then used to order the UEs based on a new ratio $V_k/c_k$, as well as to allocate the bandwidth. The DQS is presented in Algorithm\ref{alg:schedule}. \begin{algorithm}[h]
\hspace*{\algorithmicindent} \textbf{Input}  { A queue of $\it{K}$ UEs waiting for scheduling
total available bandwidth $\it{B}$;}\\
	\hspace*{\algorithmicindent} \textbf{Output} {Array $\alpha$, Scheduling decision of the UEs $x = [x_1,\dots,x_K]$};
	\begin{algorithmic}[1]
	 \State  // Cost Evaluation
	\For{$ k = 1 ,\dots, K$}
      \State  $r_k = 0, c = 1;$
      \While {$r_k(c)\leq r_{k,min}$ and $c\leq K $}
        \State $c \leftarrow c+1$;
        
    \State $c_k \leftarrow c$;
        \EndWhile 
    \EndFor\\
      \Return $C=[c_1 ,\dots, c_K]$  
    \State // Bandwidth Allocation  
    \State{order UEs according to their ratio ($\frac{V_k}{c_k}$) decreasingly and index them from $1$ to $K$;} 
    \For{$ k = 1 \dots K$}
         $x_k  \leftarrow 0 $; 
	\EndFor 
	\State{$A \leftarrow K$;}
	\State{$k \leftarrow 1$;}
	\While {$A \geq 0$}
	    \If{$ A- c_k \geq 0 $}
	    \State{$x_k \leftarrow 1$};
	    \State{$A \leftarrow A-c_k$};
	    \State{$\alpha_k \leftarrow c_k/K$};
	    \State{$k \leftarrow k+1$};
        \EndIf
    \EndWhile

\Return $x$ and $\alpha$ 
	\end{algorithmic}
	\caption{Data-quality based scheduling algorithm}
	\label{alg:schedule}
\end{algorithm}

%% file: 5_simulation.tex
\subsection{Experimental Setup}
We conduct the simulations on a desktop computer with a 2,6 GHz Intel i7 processor and 16 GB of memory and NVIDIA GeForce RTX 2070 Super graphic card. We used Pytorch \cite{noauthor_pytorch_nodate} for the ML library. The presented numerical results are the average of 10 independent runs in each setting.

\begin{figure}[htb]

	\begin{subfigure}[t]{0.35\textwidth}
		\includegraphics[width=1.42\linewidth]{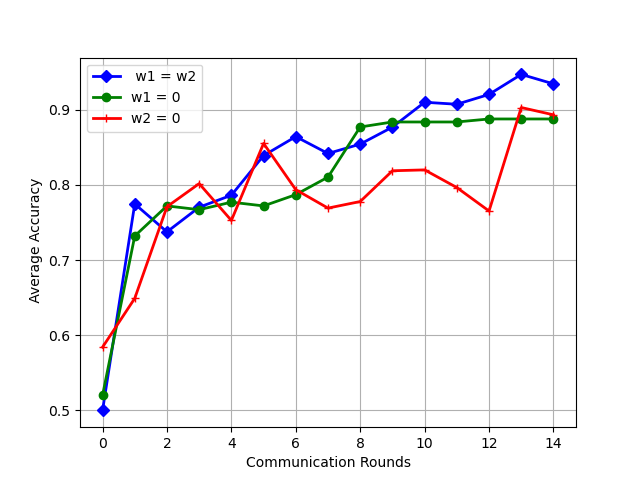}
		\caption{(6,2)}
		\label{fig:6_2}
	\end{subfigure} \hspace{18mm}
	\begin{subfigure}[t]{0.35\textwidth}
		\includegraphics[width=1.42\linewidth]{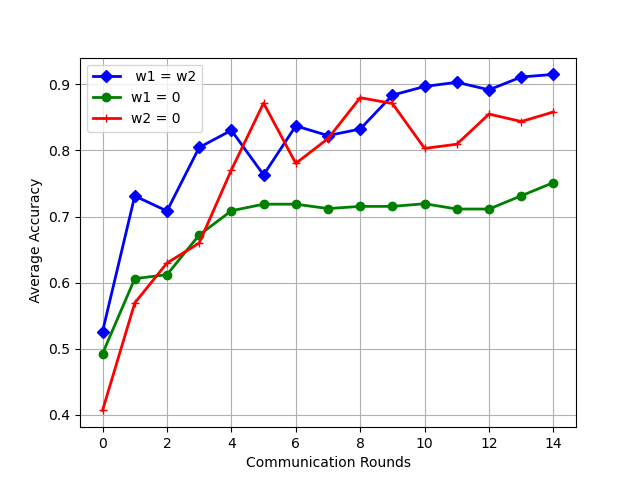}
		\caption{(8,4)}
		\label{fig:8_4}
	\end{subfigure} \hspace{6mm}

	\caption{Model accuracy depending on the targeted label following different selection strategies}
	\label{mlp_selection}
\end{figure}

\textbf{Dataset and Data poisoning:}
We  evaluate  the  efficiency  of  DQS on MNIST \cite{MNIST}, a handwritten digit images dataset. The dataset contains 50,000 training samples and 10,000 test samples labeled as 0-9.

\textit{Data distribution:}
In order to simulate non-iid and unbalanced datasets, after keeping 10\% of the data for test, the data distribution we adopted for training is as follows: 
We first sort the data by digit label, then we form 1200 groups of 50 images. Each group contains images of the same digit. In the beginning of every simulation run, we randomly allocate a minimum of 1 group and a maximum of 30 groups to each of the 50 UEs. 

\textit{Data poisoning:} According to work in \cite{shen_auror_2016,cao_understanding_2019},  when launching targeted poisoning attacks (i.e., label flipping attack) on a handwritten digits classifier,  the  easiest  and  hardest  source  and target label pairs are (6,2) and (8,4), respectively. 
Accordingly, we study both targeted labels. In fact, in each run, 5 UEs chosen at random will display a malicious behaviour by poisoning data through label flipping.

\textbf{Model:}
We train a simple multi-layer perceptron (MLP) model with two fully connected layers using  the  FedAvg  algorithm~\cite{f3} over a  total of 15 rounds. This model is lightweight, thus it can be realistically trained on resource-constrained and legacy UEs.
The model size is $S = 100Ko$, and each communication round lasts $T=300s$. The training time for each simulated UE is inferred from its training time on our setup. 

\subsection{Evaluation and Discussion}

\textit{1) Data-quality evaluation}
In this part of the evaluation, we focus on the data-quality aspect independently from the wireless environment. We weight the value of dataset diversity indicator $\it{I}$ and the reputation  $\it{R}$ under the two label flipping attacks. In each round, we select 5 UEs with the highest values for $V_k$.

Since MNIST is essentially an image classification task, and we have generated highly unbalanced datasets where several UEs only have a subset of the digits, we used Gini-Simpson index to evaluate the datasets' elements diversity~\cite{taik_data-aware_2021}. We set the weights of the diversity index equally  to $\gamma_i = 1/3$ and $\eta =1$.  We evaluate the different parts of the data-quality aspects by setting different values of $\omega_1$ and $\omega_2$. 

Fig \ref{mlp_selection} shows the results for both label flipping attacks. While in both simulations, considering both aspects by setting $\omega_1 = \omega_2$ yielded better accuracy, we noticed different response to each aspect.
For the tuple (6,2), Fig \ref{fig:6_2} shows that following a selection strategy based on data-diversity can be a good strategy, while for the harder task (8,4), as shown in Fig \ref{fig:8_4}, it seldom fails to converge. 

\textit{2) DQS evaluation:}
In this part of the evaluation, we evaluate DQS under the two label flipping attacks. 
We model the cellular network as a square of side $500$ meters with one BS located in the center of the square. The $K=50$ UEs are randomly deployed inside the square following uniform distribution. The OFDMA bandwidth is $B = 1$ MHz. We set $P_k = -23dBm$ for all UEs. The  channel gains $g_k$ between UE $k$ and the BS includes large-scale pathloss and small-scale fading following Rayleigh distribution, i.e., $|g_k|^2=d_k^{-\alpha}|h_k|^2$ where $h_k$ is a Rayleigh random variable and $\alpha$ is the pathloss exponent and $d_k$ is the distance between UE $k$ and the BS.

\begin{figure}[htb]

	\begin{subfigure}[t]{0.35\textwidth}
		\includegraphics[width=1.42\linewidth]{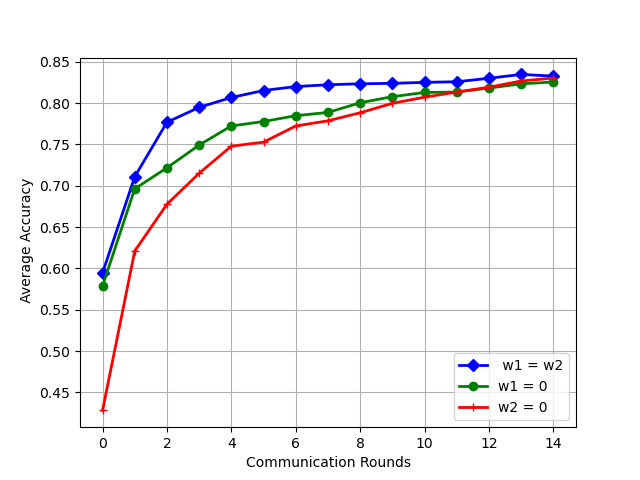}
		\caption{(6,2)}
		\label{fig:6_2_1}
	\end{subfigure} \hspace{18mm}
	\begin{subfigure}[t]{0.35\textwidth}
		\includegraphics[width=1.42\linewidth]{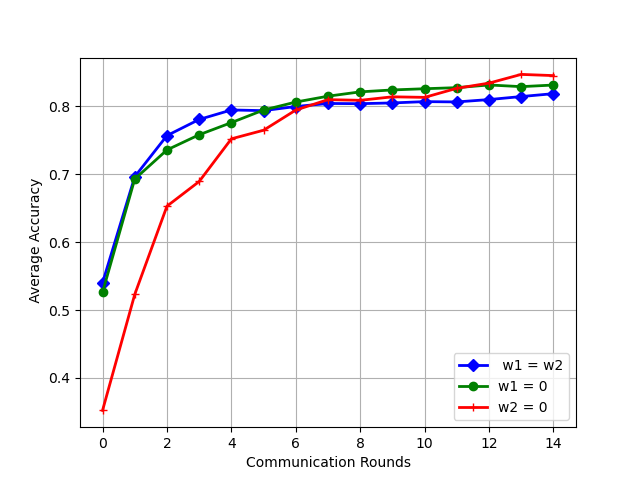}
		\caption{(8,4)}
		\label{fig:8_4_1}
	\end{subfigure} \hspace{6mm}

	\caption{Model accuracy depending on the targeted label using DQS}
	\label{mlp_scheduling}
\end{figure}

While the results using DQS in Fig.\ref{fig:6_2_1} are in concordance with the results in Fig \ref{mlp_selection},   Fig.\ref{fig:8_4_1} revealed different results. It is likely due to the varying number of participating UEs in each iteration that the results are different. Interestingly, through the results in both Fig.\ref{fig:6_2_1} 
and Fig.\ref{fig:8_4_1}, we noticed the long-term importance of the reputation aspect, on the contrast to the importance of dataset diversity in the early training rounds.  This observation becomes clear with the pair (8,4), as it makes the learning harder.
The role of reputation is far clearer at later rounds as the model becomes more sensitive to changes. As a result, an adaptive change of the weights $\omega_1$ and $\omega_2$ should be considered when using DQS.

%% file: 6_limitations.tex
Through our work on this paper, we have identified several future directions and open issues that need further investigation:
\begin{itemize}
    \item \textbf{Use-case specific measures:} While our proposed approach is general, it remains necessary to choose the adequate data-quality measures for optimal results in each use-case. 
    \item \textbf{Outliers:} Outliers in the context of FEEL  might be wrongfully considered as malicious. A tractable solution can be combining the proposed approach with clustering techniques.  
    \item \textbf{Other poisoning attacks:} Our proposed algorithm can be extended to handle other attacks such as model poisoning and multi-task poisoning attacks.
\end{itemize}

%% file: 7_conclusion.tex
FEEL is the future of distributed training in wireless edge networks by virtue of its privacy preserving aspect. In this paper, we investigated  scheduling of participant UEs in the collaborative training based on data-quality aspects. To this end, we proposed a data-quality based scheduling (DQS) algorithm for FEEL. DQS prioritizes devices with rich and diverse datasets, and punishes devices with poisoned datasets. We first defined the different components of the learning algorithm and the data-quality evaluation, namely dataset diversity and UE reputation.  Then, we formulated a joint UE selection and  bandwidth allocation problem, which we proved to be NP-hard. We presented our DQS algorithm for FEEL, which solves the problem in a greedy fashion. Finally, we evaluated the algorithm in different data poisoning scenarios, which showed the importance of data-quality evaluation components in scheduling UEs.
In the future, we will enhance our proposed algorithm to handle other attacks such as model poisoning and multi-task poisoning attacks. Furthermore, we will run evaluations on larger scale by testing on other datasets and larger number of UEs. 

%% file: bare_jrnl.bbl
\begin{thebibliography}{10}
\providecommand{\url}[1]{#1}
\csname url@samestyle\endcsname
\providecommand{\newblock}{\relax}
\providecommand{\bibinfo}[2]{#2}
\providecommand{\BIBentrySTDinterwordspacing}{\spaceskip=0pt\relax}
\providecommand{\BIBentryALTinterwordstretchfactor}{4}
\providecommand{\BIBentryALTinterwordspacing}{\spaceskip=\fontdimen2\font plus
\BIBentryALTinterwordstretchfactor\fontdimen3\font minus
  \fontdimen4\font\relax}
\providecommand{\BIBforeignlanguage}[2]{{%
\expandafter\ifx\csname l@#1\endcsname\relax
\typeout{** WARNING: IEEEtran.bst: No hyphenation pattern has been}%
\typeout{** loaded for the language `#1'. Using the pattern for}%
\typeout{** the default language instead.}%
\else
\language=\csname l@#1\endcsname
\fi
#2}}
\providecommand{\BIBdecl}{\relax}
\BIBdecl

\bibitem{r8}
\BIBentryALTinterwordspacing
J.~Konečný, H.~B. McMahan, D.~Ramage, and P.~Richtárik, ``Federated
  {Optimization}: {Distributed} {Machine} {Learning} for {On}-{Device}
  {Intelligence},'' \emph{arXiv:1610.02527 [cs]}, Oct. 2016. [Online].
  Available: \url{http://arxiv.org/abs/1610.02527}
\BIBentrySTDinterwordspacing

\bibitem{tak_federated_2021}
A.~Tak and S.~Cherkaoui, ``Federated {Edge} {Learning}: {Design} {Issues} and
  {Challenges},'' \emph{IEEE Network}, vol.~35, no.~2, pp. 252--258, Mar. 2021.

\bibitem{bonawitz_towards_2019}
\BIBentryALTinterwordspacing
K.~Bonawitz, H.~Eichner, W.~Grieskamp, D.~Huba, A.~Ingerman, V.~Ivanov,
  C.~Kiddon, J.~Konečný, S.~Mazzocchi, H.~B. McMahan, T.~Van~Overveldt,
  D.~Petrou, D.~Ramage, and J.~Roselander, ``Towards {Federated} {Learning} at
  {Scale}: {System} {Design},'' \emph{arXiv:1902.01046 [cs, stat]}, Mar. 2019,
  arXiv: 1902.01046. [Online]. Available: \url{http://arxiv.org/abs/1902.01046}
\BIBentrySTDinterwordspacing

\bibitem{moudoud1}
H.~Moudoud, S.~Cherkaoui, and L.~Khoukhi, ``An {IoT} {Blockchain}
  {Architecture} {Using} {Oracles} and {Smart} {Contracts}: the {Use}-{Case} of
  a {Food} {Supply} {Chain},'' in \emph{2019 {IEEE} 30th {Annual}
  {International} {Symposium} on {Personal}, {Indoor} and {Mobile} {Radio}
  {Communications} ({PIMRC})}, Sep. 2019, pp. 1--6.

\bibitem{abouaomar_resource_2021}
A.~Abouaomar, S.~Cherkaoui, Z.~Mlika, and A.~Kobbane, ``Resource {Provisioning}
  in {Edge} {Computing} for {Latency}-{Sensitive} {Applications},'' \emph{IEEE
  Internet of Things Journal}, vol.~8, no.~14, pp. 11\,088--11\,099, Jul. 2021,
  conference Name: IEEE Internet of Things Journal.

\bibitem{rh7}
\BIBentryALTinterwordspacing
T.~Nishio and R.~Yonetani, ``Client {Selection} for {Federated} {Learning} with
  {Heterogeneous} {Resources} in {Mobile} {Edge},'' \emph{ICC 2019}, pp. 1--7,
  May 2019. [Online]. Available: \url{http://arxiv.org/abs/1804.08333}
\BIBentrySTDinterwordspacing

\bibitem{rh6}
\BIBentryALTinterwordspacing
Q.~Zeng, Y.~Du, K.~K. Leung, and K.~Huang, ``Energy-{Efficient} {Radio}
  {Resource} {Allocation} for {Federated} {Edge} {Learning},''
  \emph{arXiv:1907.06040 [cs, math]}, Jul. 2019. [Online]. Available:
  \url{http://arxiv.org/abs/1907.06040}
\BIBentrySTDinterwordspacing

\bibitem{rh5}
N.~H. Tran, W.~Bao, A.~Zomaya, M.~N.~H. Nguyen, and C.~S. Hong, ``Federated
  {Learning} over {Wireless} {Networks}: {Optimization} {Model} {Design} and
  {Analysis},'' in \emph{{IEEE} {INFOCOM} 2019}, Apr. 2019, pp. 1387--1395.

\bibitem{goetz_active_2019}
\BIBentryALTinterwordspacing
J.~Goetz, K.~Malik, D.~Bui, S.~Moon, H.~Liu, and A.~Kumar, ``Active {Federated}
  {Learning},'' \emph{arXiv:1909.12641 [cs, stat]}, Sep. 2019. [Online].
  Available: \url{http://arxiv.org/abs/1909.12641}
\BIBentrySTDinterwordspacing

\bibitem{taik_data-aware_2021}
\BIBentryALTinterwordspacing
A.~Taik, Z.~Mlika, and S.~Cherkaoui, ``Data-{Aware} {Device} {Scheduling} for
  {Federated} {Edge} {Learning},'' \emph{arXiv:2102.09491 [cs]}, Feb. 2021,
  arXiv: 2102.09491. [Online]. Available: \url{http://arxiv.org/abs/2102.09491}
\BIBentrySTDinterwordspacing

\bibitem{rh4}
M.~H. ur~Rehman, K.~Salah, E.~Damiani, and D.~Svetinovic, ``Towards
  {Blockchain}-{Based} {Reputation}-{Aware} {Federated} {Learning},'' in
  \emph{{IEEE} {INFOCOM} 2020 - {IEEE} {Conference} on {Computer}
  {Communications} {Workshops} ({INFOCOM} {WKSHPS})}, Jul. 2020, pp. 183--188.

\bibitem{fedcs}
\BIBentryALTinterwordspacing
T.~Nishio and R.~Yonetani, ``Client {Selection} for {Federated} {Learning} with
  {Heterogeneous} {Resources} in {Mobile} {Edge},'' \emph{ICC 2019}, pp. 1--7,
  May 2019. [Online]. Available: \url{http://arxiv.org/abs/1804.08333}
\BIBentrySTDinterwordspacing

\bibitem{rh8}
W.~Xia, W.~Wen, K.-K. Wong, T.~Q. Quek, J.~Zhang, and H.~Zhu,
  ``Federated-{Learning}-{Based} {Client} {Scheduling} for {Low}-{Latency}
  {Wireless} {Communications},'' \emph{IEEE Wireless Communications}, vol.~28,
  no.~2, pp. 32--38, Apr. 2021.

\bibitem{rh9}
\BIBentryALTinterwordspacing
X.~Ma, H.~Sun, and R.~Q. Hu, ``Scheduling {Policy} and {Power} {Allocation} for
  {Federated} {Learning} in {NOMA} {Based} {MEC},'' \emph{arXiv:2006.13044 [cs,
  eess, stat]}, Jun. 2020. [Online]. Available:
  \url{http://arxiv.org/abs/2006.13044}
\BIBentrySTDinterwordspacing

\bibitem{zeng_energy-efficient_2019}
\BIBentryALTinterwordspacing
Q.~Zeng, Y.~Du, K.~K. Leung, and K.~Huang, ``Energy-{Efficient} {Radio}
  {Resource} {Allocation} for {Federated} {Edge} {Learning},''
  \emph{arXiv:1907.06040 [cs, math]}, Jul. 2019. [Online]. Available:
  \url{http://arxiv.org/abs/1907.06040}
\BIBentrySTDinterwordspacing

\bibitem{yang_age-based_2020}
H.~H. Yang, A.~Arafa, T.~Q.~S. Quek, and H.~Vincent~Poor, ``Age-{Based}
  {Scheduling} {Policy} for {Federated} {Learning} in {Mobile} {Edge}
  {Networks},'' in \emph{{ICASSP} 2020}, May 2020.

\bibitem{xia_federated-learning-based_2021}
W.~Xia, W.~Wen, K.-K. Wong, T.~Q. Quek, J.~Zhang, and H.~Zhu,
  ``Federated-{Learning}-{Based} {Client} {Scheduling} for {Low}-{Latency}
  {Wireless} {Communications},'' \emph{IEEE Wireless Communications}, vol.~28,
  no.~2, pp. 32--38, Apr. 2021, iEEE Wireless Communications.

\bibitem{rh1}
\BIBentryALTinterwordspacing
M.~Shayan, C.~Fung, C.~J.~M. Yoon, and I.~Beschastnikh, ``Biscotti: {A}
  {Ledger} for {Private} and {Secure} {Peer}-to-{Peer} {Machine} {Learning},''
  \emph{arXiv:1811.09904 [cs, stat]}, Dec. 2019, arXiv: 1811.09904. [Online].
  Available: \url{http://arxiv.org/abs/1811.09904}
\BIBentrySTDinterwordspacing

\bibitem{rh3}
\BIBentryALTinterwordspacing
Y.~Zhao, J.~Zhao, L.~Jiang, R.~Tan, D.~Niyato, Z.~Li, L.~Lyu, and Y.~Liu,
  ``Privacy-{Preserving} {Blockchain}-{Based} {Federated} {Learning} for {IoT}
  {Devices},'' \emph{arXiv:1906.10893 [cs, eess]}, Feb. 2021, arXiv:
  1906.10893. [Online]. Available: \url{http://arxiv.org/abs/1906.10893}
\BIBentrySTDinterwordspacing

\bibitem{rh2}
J.~Kang, Z.~Xiong, D.~Niyato, S.~Xie, and J.~Zhang, ``Incentive {Mechanism} for
  {Reliable} {Federated} {Learning}: {A} {Joint} {Optimization} {Approach} to
  {Combining} {Reputation} and {Contract} {Theory},'' \emph{IEEE Internet of
  Things Journal}, vol.~6, no.~6, pp. 10\,700--10\,714, Dec. 2019, iEEE
  Internet of Things Journal.

\bibitem{moudoud2}
H.~Moudoud, L.~Khoukhi, and S.~Cherkaoui, ``Prediction and {Detection} of
  {FDIA} and {DDoS} {Attacks} in {5G} {Enabled} {IoT},'' \emph{IEEE Network},
  vol.~35, no.~2, pp. 194--201, Mar. 2021.

\bibitem{cao_understanding_2019}
D.~Cao, S.~Chang, Z.~Lin, G.~Liu, and D.~Sun, ``Understanding {Distributed}
  {Poisoning} {Attack} in {Federated} {Learning},'' in \emph{2019 {IEEE} 25th
  {International} {Conference} on {Parallel} and {Distributed} {Systems}
  ({ICPADS})}, Dec. 2019, pp. 233--239.

\bibitem{bagdasaryan_how_2020}
\BIBentryALTinterwordspacing
E.~Bagdasaryan, A.~Veit, Y.~Hua, D.~Estrin, and V.~Shmatikov,
  ``\BIBforeignlanguage{en}{How {To} {Backdoor} {Federated} {Learning}},'' in
  \emph{\BIBforeignlanguage{en}{International {Conference} on {Artificial}
  {Intelligence} and {Statistics}}}.\hskip 1em plus 0.5em minus 0.4em\relax
  PMLR, Jun. 2020, pp. 2938--2948. [Online]. Available:
  \url{http://proceedings.mlr.press/v108/bagdasaryan20a.html}
\BIBentrySTDinterwordspacing

\bibitem{ye_federated_2020}
D.~Ye, R.~Yu, M.~Pan, and Z.~Han, ``Federated {Learning} in {Vehicular} {Edge}
  {Computing}: {A} {Selective} {Model} {Aggregation} {Approach},'' \emph{IEEE
  Access}, vol.~8, pp. 23\,920--23\,935, 2020.

\bibitem{lakoju_chatty_2021}
\BIBentryALTinterwordspacing
M.~Lakoju, A.~Javed, O.~Rana, P.~Burnap, S.~T. Atiba, and S.~Cherkaoui,
  ``\BIBforeignlanguage{en}{“{Chatty} {Devices}” and edge-based activity
  classification},'' \emph{\BIBforeignlanguage{en}{Discover Internet of
  Things}}, vol.~1, no.~1, p.~5, Dec. 2021. [Online]. Available:
  \url{http://link.springer.com/10.1007/s43926-021-00004-9}
\BIBentrySTDinterwordspacing

\bibitem{cormen_introduction_2009}
T.~H. Cormen, C.~E. Leiserson, R.~L. Rivest, and C.~Stein,
  \emph{\BIBforeignlanguage{en}{Introduction to {Algorithms}}}.\hskip 1em plus
  0.5em minus 0.4em\relax MIT Press, Jul. 2009.

\bibitem{noauthor_pytorch_nodate}
\BIBentryALTinterwordspacing
``\BIBforeignlanguage{en}{{PyTorch}}.'' [Online]. Available:
  \url{https://www.pytorch.org}
\BIBentrySTDinterwordspacing

\bibitem{MNIST}
Y.~Lecun, L.~Bottou, Y.~Bengio, and P.~Haffner, ``Gradient-based learning
  applied to document recognition,'' \emph{Proceedings of the IEEE}, vol.~86,
  no.~11, pp. 2278--2324, Nov. 1998, proceedings of the IEEE.

\bibitem{shen_auror_2016}
\BIBentryALTinterwordspacing
S.~Shen, S.~Tople, and P.~Saxena, ``Auror: defending against poisoning attacks
  in collaborative deep learning systems,'' in \emph{Proceedings of the 32nd
  {Annual} {Conference} on {Computer} {Security} {Applications}}, ser. {ACSAC}
  '16.\hskip 1em plus 0.5em minus 0.4em\relax New York, NY, USA: Association
  for Computing Machinery, Dec. 2016, pp. 508--519. [Online]. Available:
  \url{https://doi.org/10.1145/2991079.2991125}
\BIBentrySTDinterwordspacing

\bibitem{f3}
\BIBentryALTinterwordspacing
H.~B. {McMahan}, E.~Moore, D.~Ramage, S.~Hampson, and B.~A.~y. Arcas,
  ``Communication-efficient learning of deep networks from decentralized
  data,'' \emph{{arXiv}:1602.05629 [cs]}, 2016. [Online]. Available:
  \url{http://arxiv.org/abs/1602.05629}
\BIBentrySTDinterwordspacing

\end{thebibliography}
